\documentclass[useAMS,usenatbib]{mn2e}
\usepackage{graphicx}
\usepackage{subfigure}
\usepackage{amssymb}
\usepackage{colortbl}

\newcommand{\rmnum}[1]{\romannumeral #1}
\newcommand{\Rmnum}[1]{\expandafter\@slowromancap\romannumeral #1@}

\title[ Gravothermal oscillations ]{Gravothermal oscillations in multi-component models of star clusters}
\author[P. G. Breen and D. C. Heggie]{ Philip G. Breen$^1$\thanks{E-mail:
p.g.breen@sms.ed.ac.uk} and Douglas C. Heggie$^1$\thanks{E-mail:  d.c.heggie@ed.ac.uk} \\
$^1$ School of Mathematics and Maxwell Institute for Mathematical Sciences, University of Edinburgh, King’s Buildings, Edinburgh EH9 3JZ}

\begin{document}

\date{   \today }

\pagerange{\pageref{firstpage}--\pageref{lastpage}} \pubyear{2010}

\maketitle

\label{firstpage}

\begin{abstract}

In this paper, gravothermal oscillations are investigated in multi-component star clusters which have power law initial mass functions (IMF). For the power law IMFs, the minimum masses ($m_{min}$) were fixed and three different maximum stellar masses ($m_{max}$) were used along with different power-law exponents ($\alpha$) ranging from $0$ to $-2.35$ (Salpeter). The critical number of stars at which gravothermal oscillations first appear with increasing $N$ was found using the multi-component gas code SPEDI. The total mass ($M_{tot}$) is seen to give an approximate stability condition for power law IMFs with fixed values of $m_{max}$ and $m_{min}$ independent of $\alpha$. The value $M_{tot}/m_{max} \simeq 12000$ is shown to give an approximate stability condition which is also independent of $m_{max}$, though the critical value is somewhat higher for the steepest IMF that was studied. For appropriately chosen cases, direct N-body runs were carried out in order to check the results obtained from SPEDI. Finally, evidence of the gravothermal nature of the oscillations found in the N-body runs is presented. 
\end{abstract}

\begin{keywords}
 globular clusters: general; methods: numerical; methods: n-body simulations.
\end{keywords}


\section{Introduction}
The condition for the onset of gravothermal oscillations is best understood for the case of one-component star clusters, clusters consisting of stars of equal mass. \cite{Goodman1987} found that gravothermal oscillations first appear when the number of stars $N$ is greater than 7000.  This condition has also been confirmed with Fokker-Planck calculations \citep{Cohn_et_al1989} and by direct N-body simulations \citep{Makino1996}. However, the multi-component case is more complicated. This is due to the fact that the presence of several components introduces new dynamical processes into the system, and several additional parameters in addition to $N$. 

Even for the two-component case, which is the simplest kind of mass spectrum, the condition for the onset of gravothermal oscillations is not so simple. Two-component models can be subdivided into Spitzer stable and Spitzer unstable cases depending on whether or not the two components can achieve equipartition of kinetic energy during core collapse \citep{Spitzer}. \cite{KimLeeGood1998} studied a range of Spitzer stable two-component models. Their research supported the applicability of the Goodman stability parameter $\epsilon$  \citep[see][]{Goodman1993} as a stability criterion. \cite{breenheggie1}, whose research focused on the more general Spitzer unstable two-component case, indicated that the occurrence of gravothermal instability depends approximately on the number of stars in the heavier component. \cite{breenheggie1} also found that the critical value of $\epsilon$ depended on the parameters of the mass function (e.g. stellar mass ratio). However, by using a slightly modified version of $\epsilon$, one with a modified definition of the half mass relaxation time, they found a nearly constant critical value.

\citet{Murphyetal1990} found that the post-collapse evolution of multi-component models was stable to much higher values of $N$ than in one-component models and that the value of $N$ at which gravothermal oscillations appeared varied with different mass functions. They studied seven-component systems constructed to approximate evolved power law IMFs with masses ranging from $0.1$ to $1.2M_{\sun}$. The power law exponent that they considered ranged from $-2$ to $-4.5$. They found that gravothermal oscillations appeared when the total mass of the system ($M_{tot}$) was of order $8 \times 10^4 M_{\sun}$ \citep[see][ Figure 6]{Murphyetal1990} and that the critical value of $M_{tot}$ increased with decreasing power law exponent. They suggested that the appearance of oscillations depends on the number of heavier stars. However, this leads to the issue that in a multi-component system it is not clear what the definition of a heavy star should be (this point is discussed in Section 2).

The main aim of the present paper is to provide a theoretical understanding of the onset of gravothermal oscillations in multi-component systems. As this present paper follows on from the research of \cite{breenheggie1} it is worthwhile attempting to extend the concepts developed in that paper to the multi-component case. Although two-component systems may be realistic approximations of multi-component systems \citep{KimLee1997}, it is best to have a better understanding of gravothermal oscillations in multi-component systems as real globular clusters contain a continuous mass spectrum. What is of particular importance is the effect of varying the maximum stellar mass ($m_{max}$) on the onset of instability as this was not studied by \cite{Murphyetal1990}. 


The rest of this paper is structured as follows. In Section 2, we state the results concerning gravothermal oscillations in gaseous models. This section also contains subsections on the Goodman stability parameter and a variant which used a modified relaxation timescale. This is followed by Section 3 in which the results of N-body simulations are given. Finally, Section 4 consists of the conclusion and discussion. 



\section{ Critical Value of $N$ }\label{sec:Ncritinto}

\subsection{Results of gaseous models}\label{sec:Ncrit}
In all cases, the initial conditions used were realisations of the Plummer model \citep{Plummer1911, HeggieHut2003}. The initial velocity dispersion of all components and the initial ratio of density of all components were equal at all locations. The initial conditions were constructed in order to approximate a continuous power law IMF with different exponents $\alpha = -0.0, -0.65, -1.3, -1.65, -2.00$ and $-2.35$. The multi-component gas code SPEDI\footnote{ SPEDI is a multi-component gas code which was initially based on a formulation by \cite{Louisspurzem} and was subsequently further developed by \cite{spurzemtakahashi}. Further information regarding SPEDI is available at http://www.ari.uni-heidelberg.de/gaseous-model/. } was used for all gaseous models in the present paper. The power law IMFs were approximated by dividing the complete mass range into equal logarithmic steps. Alternative methods of discretization were also tried for certain cases in order to confirm the validity of the results, such as approximating the IMF using equal total masses in each of the components. The ranges of stellar masses $(m_{max},m_{min})$ used in this paper are $(1.0,0.1)$, $(2.0,0.1)$ and $(3.0,0.1)$. The reason why higher values of $m_{max}$ were not used is that it is customary to suppose that a cluster would be largely depleted in heavier stars by the time gravothermal oscillations manifest \citep{KimLeeGood1998}. In Sec \ref{sec:conanddis}, however, we briefly discuss a possible exception.

The critical value of $N$ ($N_{crit}$) at which oscillations in the central density ($\rho_c$) first appeared (as $N$ increased) was determined (correct to ten percent). The obtained values of $N_{crit}$ in units of $10^4$ are given in Table \ref{table:tab1}.

\begin{table}
\begin{center}
\caption{Critical value of $N$ ($N_{crit}$) in units of $10^4$. The values of $N_{crit}$ in brackets were obtained using  5-component models, while all other values were obtained using 10-component models. The value of $N_{ef}$ for the case $\alpha = -2.35$ and extreme masses $(3,0.1)$ could not be obtained with a 10-component model due to numerical difficulties.}
\begin{tabular}{c| c c c c c c}
  $\!\!\!\!\!(m_{max},m_{min})\backslash\alpha\!\!\!$                        & \bf{0} & \bf{-0.65} & \bf{-1.3} & \bf{-1.65}  & \bf{-2.0} & \bf{-2.35} \\ \hline
\bf{(3,0.1)}             &  2.0   &  2.8       &  5.2      & 8.0           & 12.0       & \hspace{2 mm}--\hspace{2 mm}    \\ 
    &        &        &            &           &       $(10.0)$        &  $(15.7)$          \\ 
 \bf{(2,0.1)}               &  2.0   &  2.6       &  4.5      & 7.0         & 8.0        & 10.0       \\ 
   &        &              &         &            &           &  $(10.0)$ \\ 
 \bf{(1,0.1)}               &  2.0   &  2.5       &  3.5      & 4.5         & 6.0        & 8.0        
\label{table:tab1}
\end{tabular}
\end{center}
\end{table}

\subsection{Interpretation of results}\label{sec:lsm}

Guided by the results of \cite{Murphyetal1990}, we first consider the values of $M_{tot}$ at $N_{crit}$ ($M_{crit}$). The values of $M_{crit}$, for the models considered in the present paper, are given in Table \ref{table:mcrit}. The values of $M_{crit}$ in Table \ref{table:mcrit} are approximately the same for fixed $m_{max}$ and the values vary much less with $\alpha$ than $N_{crit}$. Thus the conclusion of \cite{Murphyetal1990}, who considered only evolved IMFs with fixed $m_{max}$, appears also to apply to pure power law IMFs with fixed $m_{max}$. What has been added in the study in the present paper is that the value of $M_{crit}$ also has a strong dependence on $m_{max}$. 

We can compare the dependence on $m_{max}$ in Table \ref{table:mcrit} with the results of the two-component models of \cite{breenheggie1} if we fix the stellar mass of the light component in that earlier paper. This is done in Appendix \ref{A} (see Table \ref{table:aptab2}). In Table \ref{table:aptab2} there is a clear trend of increasing $M_{crit}$ with increased stellar mass of the heavy component for fixed total mass ratio (i.e. moving up through one column of Table \ref{table:aptab2}). Therefore the trend of increasing $M_{crit}$ with increasing stellar mass range (or stellar mass ratio), for fixed $m_{min}$, seems to be a common feature of multi-component systems. It is also worth noting that for two-component systems with fixed stellar mass ratio (see Appendix \ref{A} Table \ref{table:aptab2} where we consider values along a given row) there is a trend of increasing $M_{crit}$ with decreasing total mass ratio. As decreasing total mass ratio for two-component systems is the analogue of decreasing $\alpha$, these systems have the same stability trends as the multi-component systems in Table \ref{table:mcrit}. Now we will attempt to extend the interpretation of \cite{breenheggie1} from two-component to multi-component models in a way which also accounts for the dependence on $m_{max}$. 

\begin{table}
\begin{center}
\caption{Critical value of $M$ ($M_{crit}$) in units of $10^4$}
\begin{tabular}{ c  |c c c c cc}
 $\!\!\!\!\!(m_{max},m_{min})\backslash\alpha\!\!\!$      & \bf{0}    & \bf{-0.65}  & \bf{-1.3} &\bf{-1.65} &\bf{-2.00}&  \bf{-2.35} \\  \hline
                 \bf{(3,0.1)} &  3.1    &  3.1      &  3.4    & 3.8  &  4.2 & $(4.3)$  \\ 
                 \bf{(2,0.1)} &  2.1    &  2.0      &  2.3    & 2.8  &  2.5 & 2.6       \\
                 \bf{(1,0.1)} &  1.1    &  1.1      &  1.2    & 1.3  &  1.5 & 1.8       
\label{table:mcrit}
\end{tabular}
\end{center}
\end{table}

\cite{breenheggie1} first argued that for the two-component case, the dynamics of the system were dominated by the heavier component. The reasoning behind this emphasis on $N_2$ is that the heavier component concentrates within the central region where it behaves like a one-component system deep in the potential well of the low mass stars, and it can exhibit gravothermal instability like the central part of a one-component system\footnote{ Note that, in this picture, the gravothermal instability  of the system is essentially confined to the massive stars; it is in the centrally concentrated massive stars that the temperature inversions occur which drive the expansion phase of the gravothermal oscillations. Around $N=N_{crit}$, the light stars always exhibit a normal temperature gradient, and if their heat capacity is negative this does not lead to instability.}. \cite{breenheggie1} showed that $N_2$, the number of heavy stars, does indeed  provide an approximate criterion for the onset of instability, and found the critical value of $N_2$ to be of order $2000$. For the case of multi-component models, however, it is unclear if and how the system can be divided into a heavy and a light component. Nonetheless, it may still be expected that the heavier stars may be more important to the dynamics of the system and to the onset of gravothermal oscillations. 

\cite{breenheggie1} then gave an alternative measure of the importance of the heavy component, in terms of what they called the ``effective" number of heavy stars, and this is a concept that is more readily adapted to the case of several components. \cite{breenheggie1} argued that, as the light component acts as a kind of container for the heavy component, it was the overall mass of this container (i.e. the total mass in the light component) that was the important factor and not the stellar mass of the light component. Therefore it could be replaced by an equal mass of stars of the massive component, giving rise to the idea of the effective number of stars $N_{ef}$. This was defined as follows \begin{center}
\begin{equation}\label{eq:Nef}
\quad
\qquad \qquad \qquad   N_{ef} = \frac{M_1 + M_2}{m_2}.
\end{equation}
\end{center}
They also defined a modified half-mass relaxation time scale $(t_{ef,rh})$ by using $N_{ef}$ in place of $N$ in the standard formula for the half-mass relaxation time.  They used this effective relaxation time to modify and improve upon a stability criterion suggested by \cite{Goodman1993}; see Section \ref{sec:eps} of the present paper. It is worth pointing out that $N_{ef}$ itself can be used as an approximate stability condition for the two-component models of \cite{breenheggie1} (see Appendix \ref{A}). It is this form of condition for the stability of two-component systems which we shall attempt to generalise to multi-component systems.

Again it is not immediately clear what the most appropriate extension of $N_{ef}$ to the multi-component case should be, as the appropriate definition of heavy stars is not clear. However, the result is much less sensitive to our choice than the number of heavy stars, $N_2$. One simple approximate way to define a heavy star in this context is simply any star with a stellar mass of $\approx m_{max}$. This would lead to the definition of $N_{ef}$ as $\frac{M_{tot}}{m_{max}}$ for the multi-component model and no change in the definition for a two-component model (equation (\ref{eq:Nef})). This value is given in Table \ref{table:mcritNef}, for the multi-component models considered in this paper. We find that there is much less variation among the critical values of $N_{ef}$ than for $M_{crit}$, especially for varying $m_{max}$. Nevertheless, the same trend of increasing $M_{crit}$ with decreasing $\alpha$ as seen in Table \ref{table:mcrit} is still present in Table \ref{table:mcritNef} in the form of increasing $N_{ef}$.

\begin{table}
\begin{center}
\caption{Critical value of $N_{ef}$ in units of $10^4$.  }
\begin{tabular}{ c  | c c c c c c}
  $\!\!\!\!\!(m_{max},m_{min})\backslash\alpha\!\!\!$       & \bf{0}    & \bf{-0.65}  & \bf{-1.3} &   \bf{-1.65} &\bf{-2.00}& \bf{-2.35} \\ \hline 
  \bf{(3,0.1)} &  1.1    &  1.0      &  1.1    & 1.3        &  1.4  &    $(1.4)$          \\ 
  \bf{(2,0.1)} &  1.1    &  1.0      &  1.2    & 1.4        &  1.3   &    1.3      \\ 
  \bf{(1,0.1)} &  1.1    &  1.1      &  1.2    & 1.3        &  1.5   &    1.8       
\label{table:mcritNef}
\end{tabular}
\end{center}
\end{table}

Now we will discuss a possible explanation for the increase in the critical value of $N_{ef}$ with decreasing $\alpha$ in Table \ref{table:mcritNef}. The idea behind $N_{ef}$ is that a multi-component system behaves in approximately the same way as a single-component system consisting of $N_{ef}$ stars of stellar mass $m_{max}$. As systems with higher $\alpha$ have more stars with stellar mass $\approx m_{max}$ than systems with lower $\alpha$, we may expect that the approximation with the one-component system is better for higher $\alpha$ than for lower $\alpha$. Therefore, it is not surprising that in Table \ref{table:mcritNef} it is the systems with $\alpha=0$ that have the closest critical values of $N_{ef}$ to the critical value of $N$ for a one-component system, i.e. about $7000$.    

We can take our discussion further by considering systems with fixed $N_{ef}$. For systems with fixed $N_{ef}$, as $\alpha$ decreases there is an increasing number of light stars (stars with stellar mass not $\approx m_{max}$). As the number of light stars increases and the number of heavy stars decreases the two-body relaxation time increases, as it becomes increasingly dominated by the light component. According to H\'{e}non's principle \citep{Henon2} the rate of energy generation in the system is regulated by two-body relaxation, and therefore there is a lower rate of energy generation as $\alpha$ decreases. Regardless of the value of $\alpha$, the average mass in the core is approximately $m_{max}$, and the lower rate of energy generation can be met by a core of lower density. Thus for lower alpha there is a smaller density contrast between the core and the mean density in the system. This would imply that the stability would increase with decreasing $\alpha$, as we indeed see.

\subsection{Goodman stability parameter}\label{sec:eps}
 
\cite{Goodman1993} suggested the use of the quantity
\begin{center}
\begin{equation}\label{eq:eps1}
\quad
\qquad \qquad \qquad   \epsilon \equiv \frac{E_{tot}/t_{rh} }{E_c /t_{rc}}
\end{equation}
\end{center}
as a stability indicator, where $\log_{10}{\epsilon} \sim -2$ is the stability limit below which the cluster becomes unstable. Here $E_{tot}$ is the total energy, $E_c$ is the energy of the core, $t_{rc}$ is the core relaxation time and $t_{rh}$ is the half mass relaxation time. A condition of this type has been supported for two-component models by \cite{KimLeeGood1998} who studied Spitzer stable models using a Fokker-Planck code and by \cite{breenheggie1} who studied Spitzer unstable models using a gas code. However, \cite{breenheggie1} also introduced a modified definition of $\epsilon$ (see below) because the definition given in equation (\ref{eq:eps1}) was found to yield a critical value which varied with total mass ratio and stellar mass ratio. Also, the critical value at which the instability appears is somewhat different in \cite{breenheggie1} from that in \cite{KimLeeGood1998}.

For the multi-component models studied in the present paper, the values of $\log_{10} \epsilon$ are given in Table \ref{table:goodmanepsilon1}. The values of $\log_{10} \epsilon$ (based on the original definition, i.e. equation (\ref{eq:eps1})) range from $-2.26$ to $-2.56$ and there is a decreasing trend with decreasing $\alpha$.

\begin{table}
\begin{center}
\caption{ Critical value of $\log_{10} \epsilon $ }
\begin{tabular}{ l |c c c ccl}
   $\!\!\!\!\!\!(m_{max},m_{min})\backslash\alpha\!\!\!\!$             & \bf{0} $\!$ & \bf{-0.65}$\!$ & \bf{-1.3}  &\bf{-1.65}&\bf{-2.00} &  \bf{-2.35} \\ \hline
  \bf{(3,0.1)} & -2.26 $\!$  & -2.38$\!$ & -2.54 & -2.61 & -2.57 &   (-2.62) \\ 
\bf{(2,0.1)} & -2.26 $\!$ & -2.32$\!$ & -2.48 & -2.53 & -2.58 & -2.56 \\ 
  \bf{(1,0.1)} & -2.26 $\!$ & -2.30$\!$ & -2.39 & -2.40 & -2.43 & -2.42   
\label{table:goodmanepsilon1}
\end{tabular}
\end{center}
\end{table}

In equation (\ref{eq:eps1}), $t_{rc}$ and $t_{rh}$ are defined by
\begin{equation}\label{eq:trc}
  t_{rc} =
\frac{0.34  \bar{\sigma}^3_{c} }{G^2 \bar{m_c} \rho_{c}  \ln { \Lambda}}
\end{equation}
and
\begin{equation}\label{eq:trheq}
 t_{rh} =
\frac{0.138N^{\frac{1}{2}}r_h^{^{\frac{3}{2}}}}{(G\bar{m})^{\frac{1}{2}} \ln { \Lambda}}.
\end{equation}
where $\bar{m}_c$ and $\bar{\sigma}_c$ are the mass density weighted averages over all the components in the core. However, the definition of $t_{rh}$ does not take into account the mass spectrum. \cite{breenheggie1} have shown that for two-component models, modifying the definition of $t_{rh}$ to take into account the mass spectrum leads to an improved stability condition. As has already been shown in Section \ref{sec:Ncrit}, we can construct a value $N_{ef}$ which provides an approximate stability condition. Now we will use this value to define a new half mass relaxation time defined as
\begin{equation}\label{eq:trheq2}
 t_{ef,rh} =
\frac{0.138N_{ef}^{\frac{1}{2}}r_h^{^{\frac{3}{2}}}}{(Gm_{max})^{\frac{1}{2}} \ln { \Lambda}}.
\end{equation}
We use this in place of $t_{rh}$ in equation (\ref{eq:eps1}) to define the new stability parameter $\epsilon_2$ as in \cite{breenheggie1}.  The values of this parameter are given in Table \ref{table:goodmanepsilon2}. The variation of $\log_{10} \epsilon_2 $ in Table \ref{table:goodmanepsilon2} is of comparable magnitude to the variation of $\log_{10} \epsilon$ in Table \ref{table:goodmanepsilon1}, but the values in Table \ref{table:goodmanepsilon2} are more consistent with that of a one-component model ($\log_{10} \epsilon = \log_{10} \epsilon_2  = -2$) than those in Table \ref{table:goodmanepsilon1}.

For the one-component model the definitions of $\epsilon$ and $\epsilon_2$ are identical. Also, it is worth noting that, for two-component systems with extremely small amounts of the heavy component (relative to $M_{tot}$), $t_{ef,rh}$ as defined in equation (\ref{eq:trheq2}) is not a suitable approximation to the relaxation time. This is the case for the models considered by \cite{KimLeeGood1998}, and so the extension to $\epsilon_2$ would not have been necessary or useful in the context of their paper.

The logarithm of the Goodman stability parameter (or our somewhat more consistent modified version) seems to have a particular value approximately $-2$ at the stability boundary (\cite{KimLeeGood1998}, \cite{breenheggie1} and the present paper). 
However, it is not known if $\epsilon$ (or $\epsilon_2$) can be predicted for a particular IMF without carrying out numerical simulations, which limits its usefulness. In contrast $N_{ef}$ has the advantage that it can be easily calculated before carrying out numerical simulations, and also provides an approximate indication of the stability boundary.

\begin{table}
\begin{center}
\caption{ Critical value of $\log_{10} \epsilon_2 $ }
\begin{tabular}{ l  |c c c ccl}
          $\!\!\!\!\!(m_{max},m_{min})\backslash\alpha\!\!\!$         & \bf{0$\!$}  & \bf{-0.65$\!$}   & \bf{-1.3}$\!$ &\bf{-1.65}$\!$ & \bf{-2.00}$\!$ &  \bf{-2.35}$\!$ \\ \hline%
     \bf{(3,0.1)} &  -2.06  &  -2.02       & -1.99     & -1.94   &  -1.75  & (-1.73) \\ 
     \bf{(2,0.1)} &  -2.04  &  -1.99       & -2.00     & -1.95  &  -1.75  & -1.81  \\ 
     \bf{(1,0.1)} &  -2.05  &  -2.01       & -2.01     & -1.96  &  -1.94  & -1.87   \\ 
\label{table:goodmanepsilon2}
\end{tabular}    
\end{center}
\end{table}
%

\section{ Direct N-body Simulations }\label{sec:dnbody}


In order to validate the results obtained from the gas code SPEDI a series of $N$-body runs were carried out. The direct $N$-body simulations in the present paper were conducted using the NBODY6 code \citep{aarseth,NitadoriAaresth2012}. As the IMF's with $\alpha=0$ in Table \ref{table:tab1} have the lowest values of $N_{crit}$ ($2.0 \times 10^{4}$), for this value of $\alpha$ $N$-body runs were carried out for the mass ranges $(1,0.1)$ and $(2,0.1)$ . The case with parameters $\alpha=-1.3$ and $(m_{max},m_{min})=(1.0,0.1)$ was also chosen because it has a higher value of $N_{crit}$ ($3.5 \times 10^{4}$) than for $\alpha=0$, although the value is still low enough to make it suitable for direct $N$-body simulations.

\begin{figure}
\subfigure{\scalebox{0.65}{\includegraphics{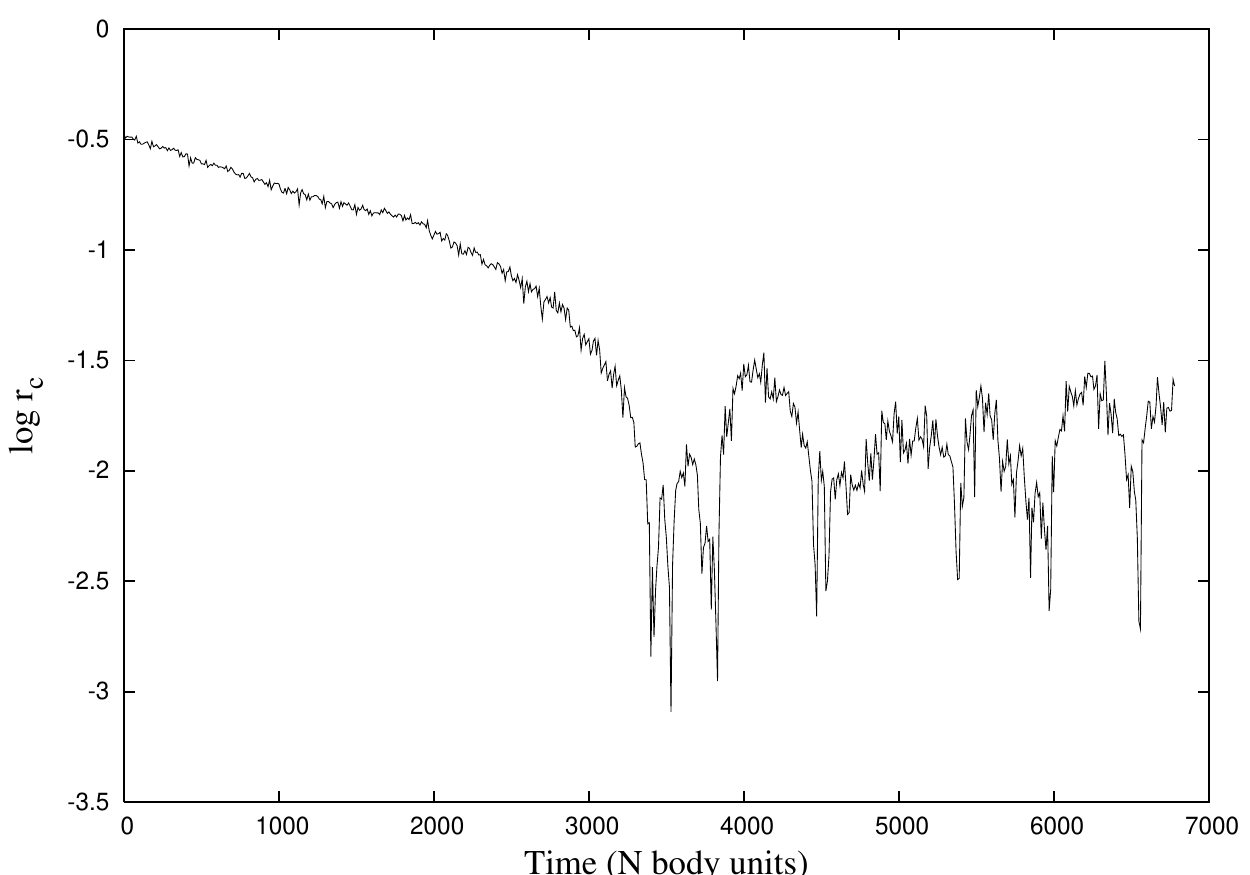}}}\quad
\subfigure{\scalebox{0.65}{\includegraphics{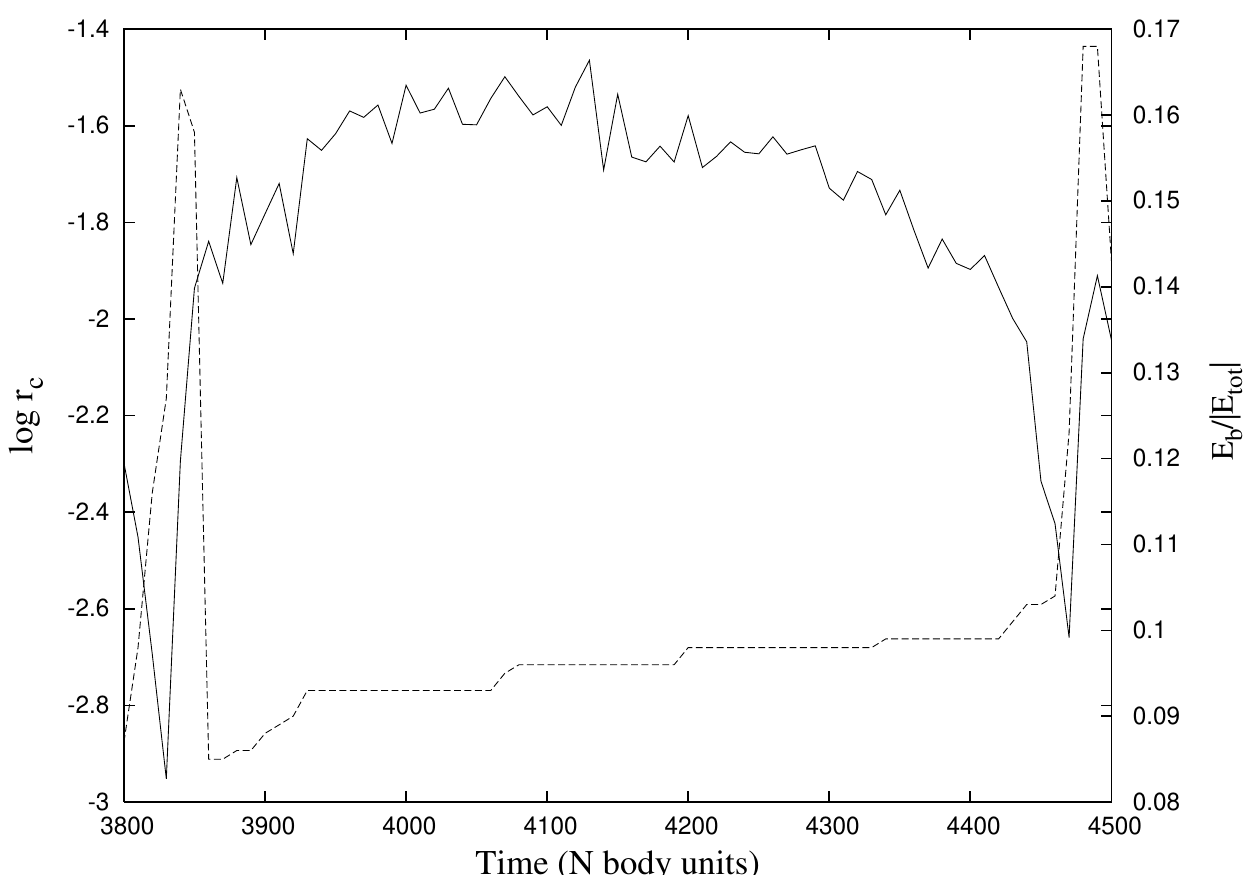}}}\quad
\subfigure{\scalebox{0.65}{\includegraphics{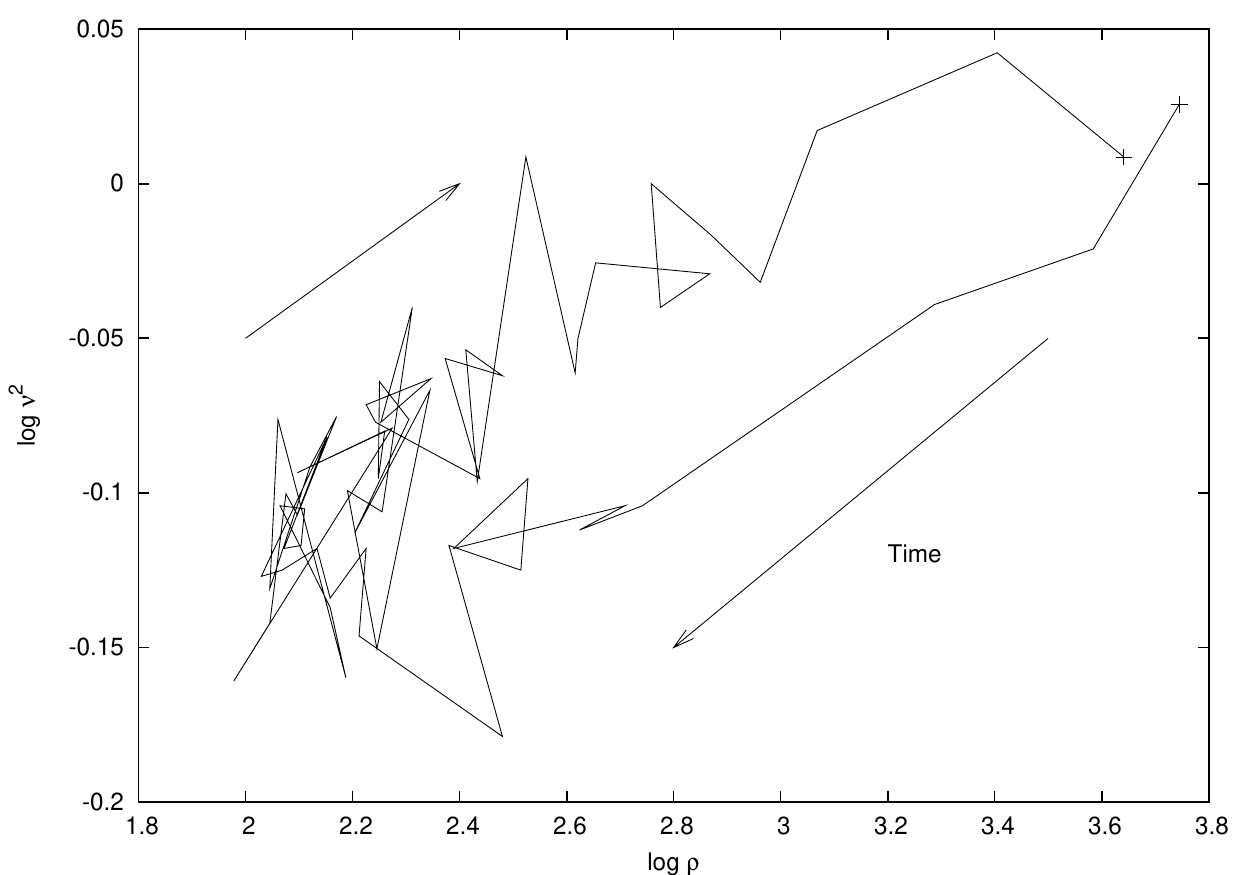}}}
\caption{ $N$-body run of a multi-component model with $N=32k$, $\alpha = 0.0$ and $(m_{max},m_{min})=(1.0,0.1)$.  $\bf{Top}$: $\log{r_{c}}$ vs time ($N$ body units) over the entire run. $\bf{Middle}$: $\log{r_{c}}$ vs time ($N$ body units)  over-plotted with the relative binding energy of binaries. The plot is over the period of a gravothermal oscillation which occurs between $3830$ and $4570$. $\bf{Bottom}$:  $\rho_c$ vs $v^2_c$, showing the gravothermal nature of the  cycle over the same time period as the middle plot.}
\label{fig:32ka0m1}
\end{figure}

\begin{figure}
\subfigure{\scalebox{0.65}{\includegraphics{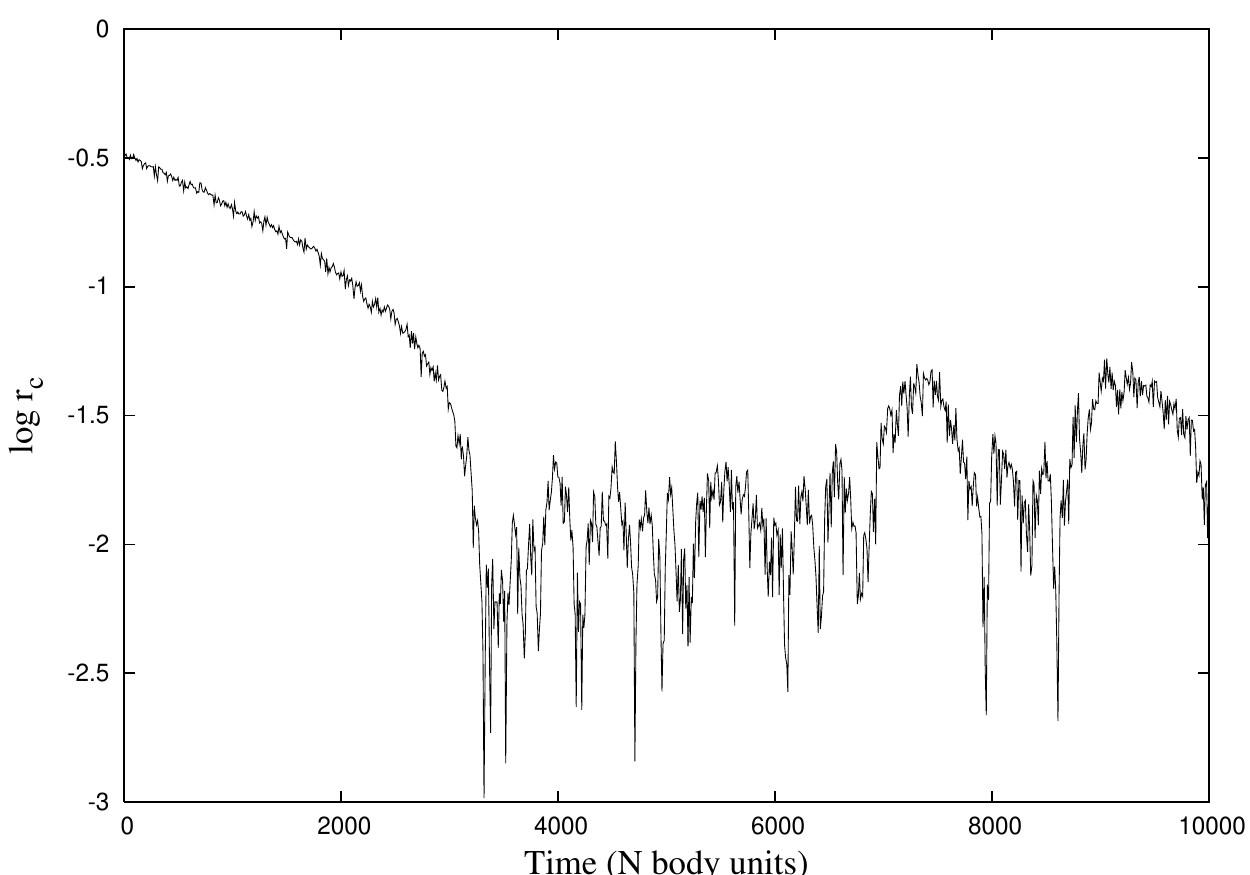}}}\quad
\subfigure{\scalebox{0.65}{\includegraphics{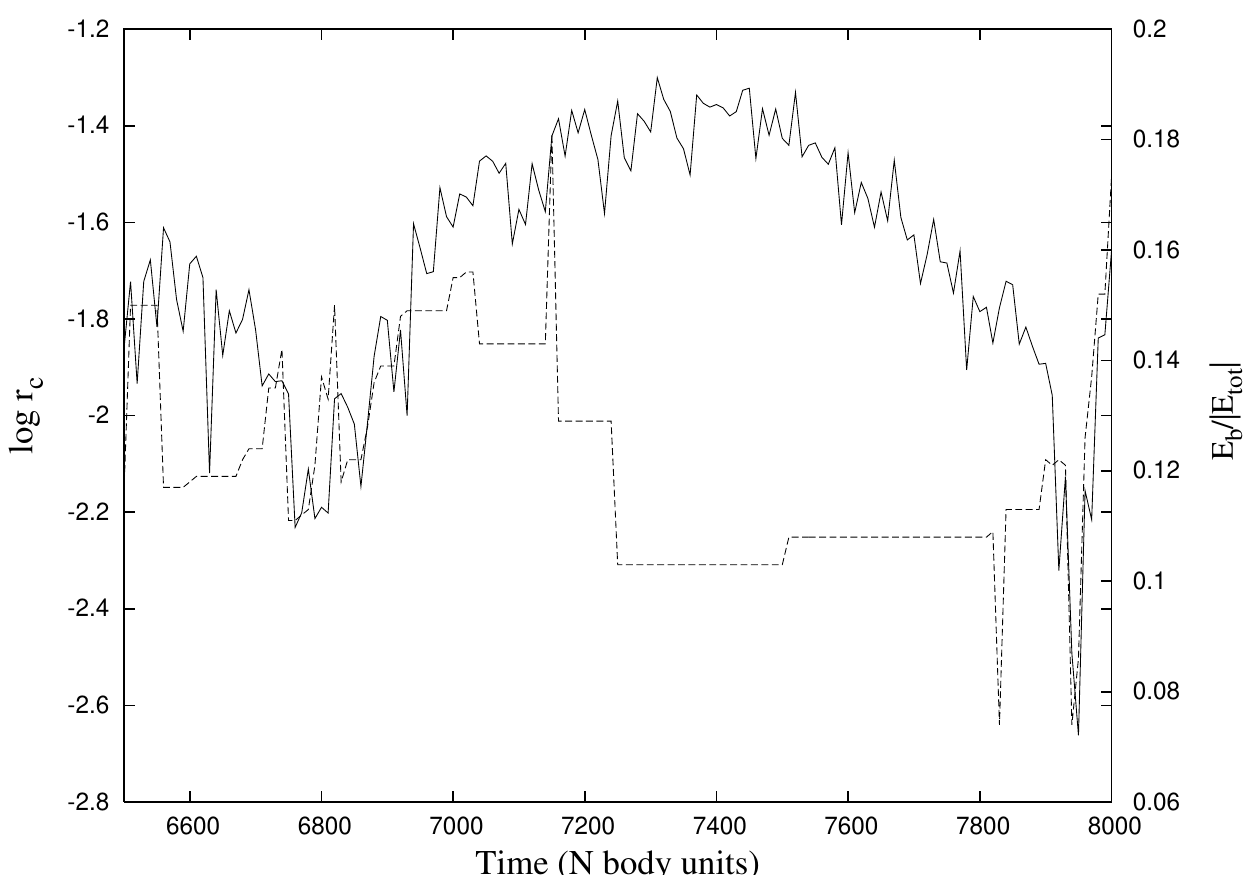}}}\quad
\subfigure{\scalebox{0.65}{\includegraphics{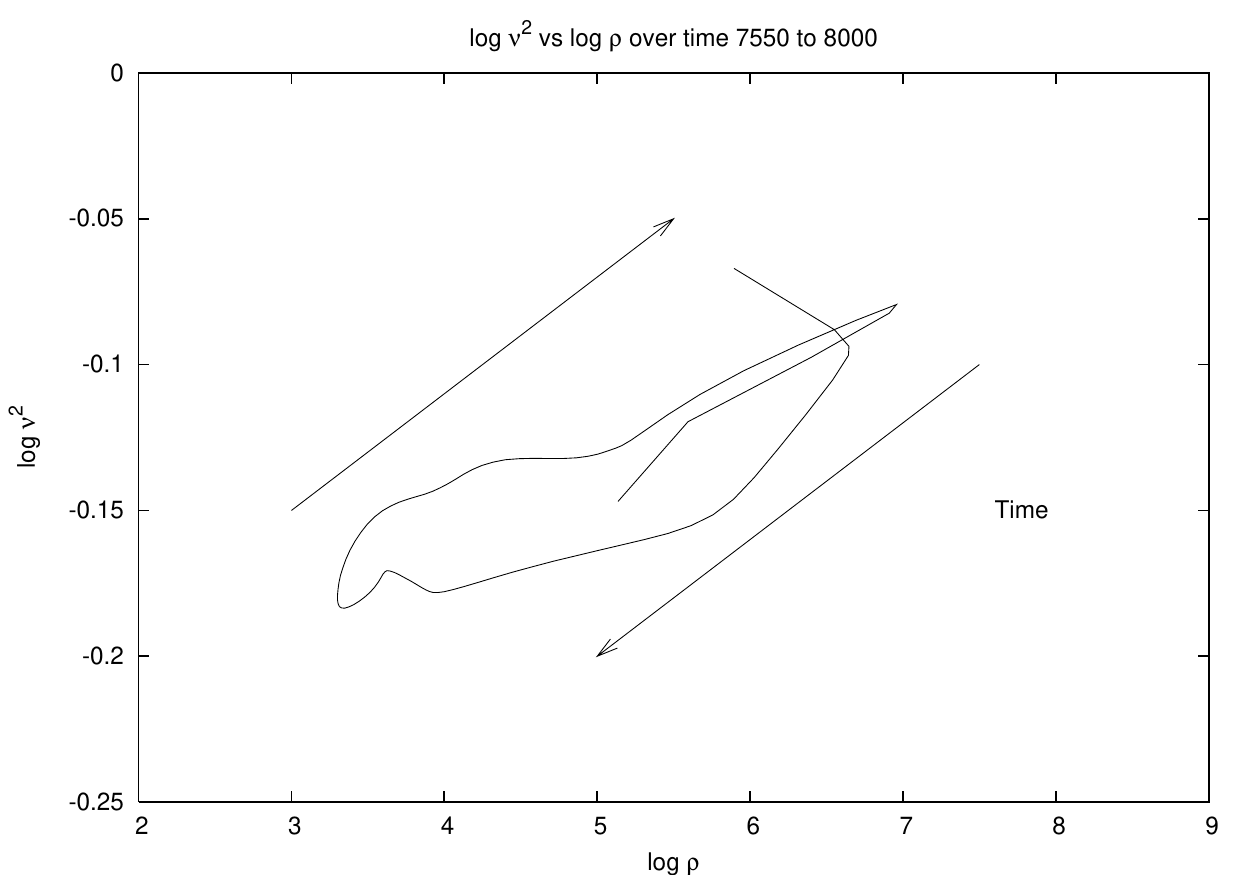}}}
\caption{ As Fig. \ref{fig:32ka0m1}, with $N=32k$ and $\alpha = 0.0$, but $(m_{max},m_{min})=(2.0,0.1)$. The velocity dispersion has been smoothed to make the cycle clearer.}
\label{fig:32ka0m2}
\end{figure}

For the case of $\alpha = 0.0$ the values of $N_{crit}$ are the same $($see Table \ref{table:tab1}$)$ regardless of the stellar mass range. For the two mass ranges chosen, there were no signs of gravothermal behaviour in the $N$-body runs with $N=8k$ or $N=16k$. The first clear sign of gravothermal behaviour occurs with $N=32k$ for both chosen mass ranges. For the stellar mass range $(1.0,0.1)$, the three panels of Fig. \ref{fig:32ka0m1} show, respectively, (\rmnum{1}) the evolution of the core radius $r_c$, (\rmnum{2}) an example of a single cycle of gravothermal oscillation in the post-collapse evolution, and (\rmnum{3}) evidence of the gravothermal nature of the oscillation for the $32k$ run. The same graphs for the $32k$ run with stellar mass range $(2.0,0.1)$ are given in Fig. \ref{fig:32ka0m2}.

\begin{figure}
\subfigure{\scalebox{0.65}{\includegraphics{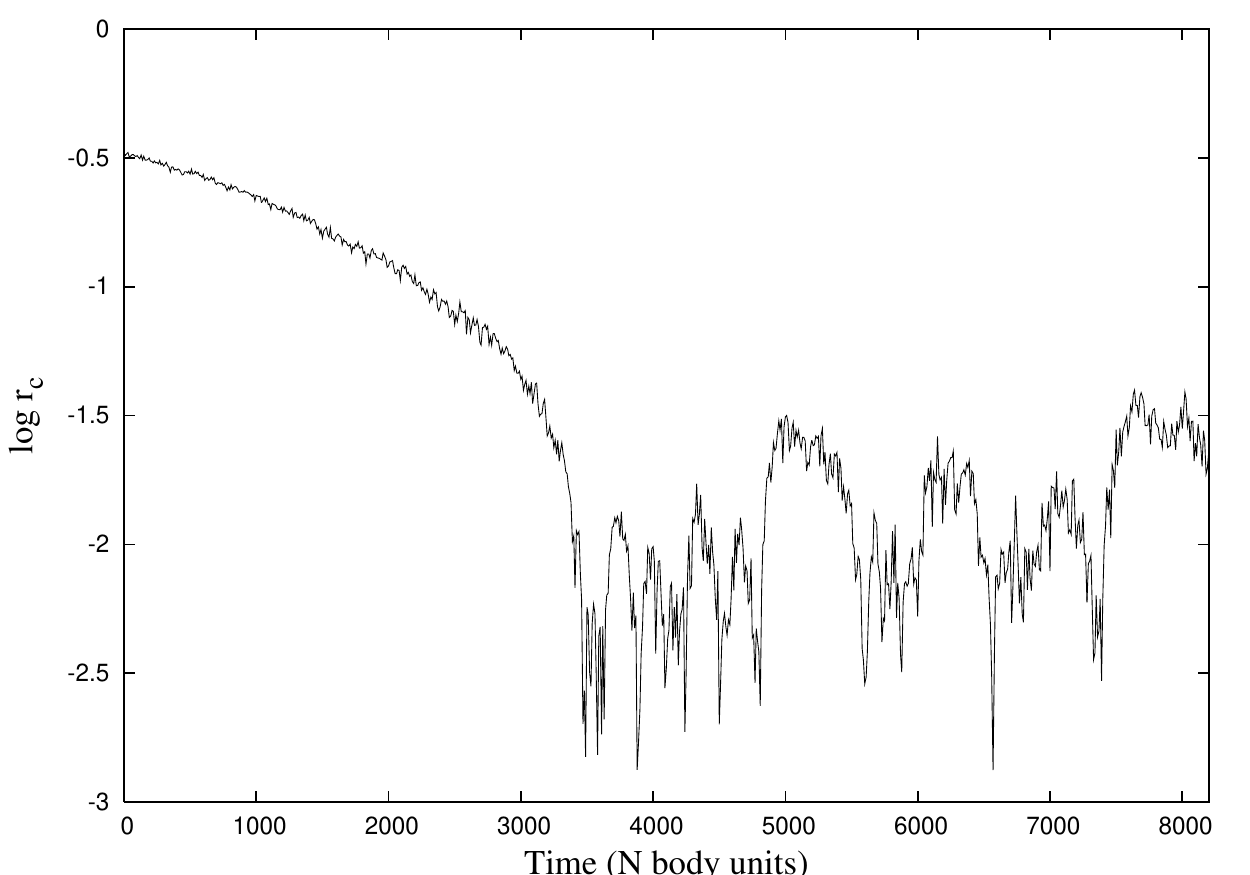}}}\quad
\subfigure{\scalebox{0.65}{\includegraphics{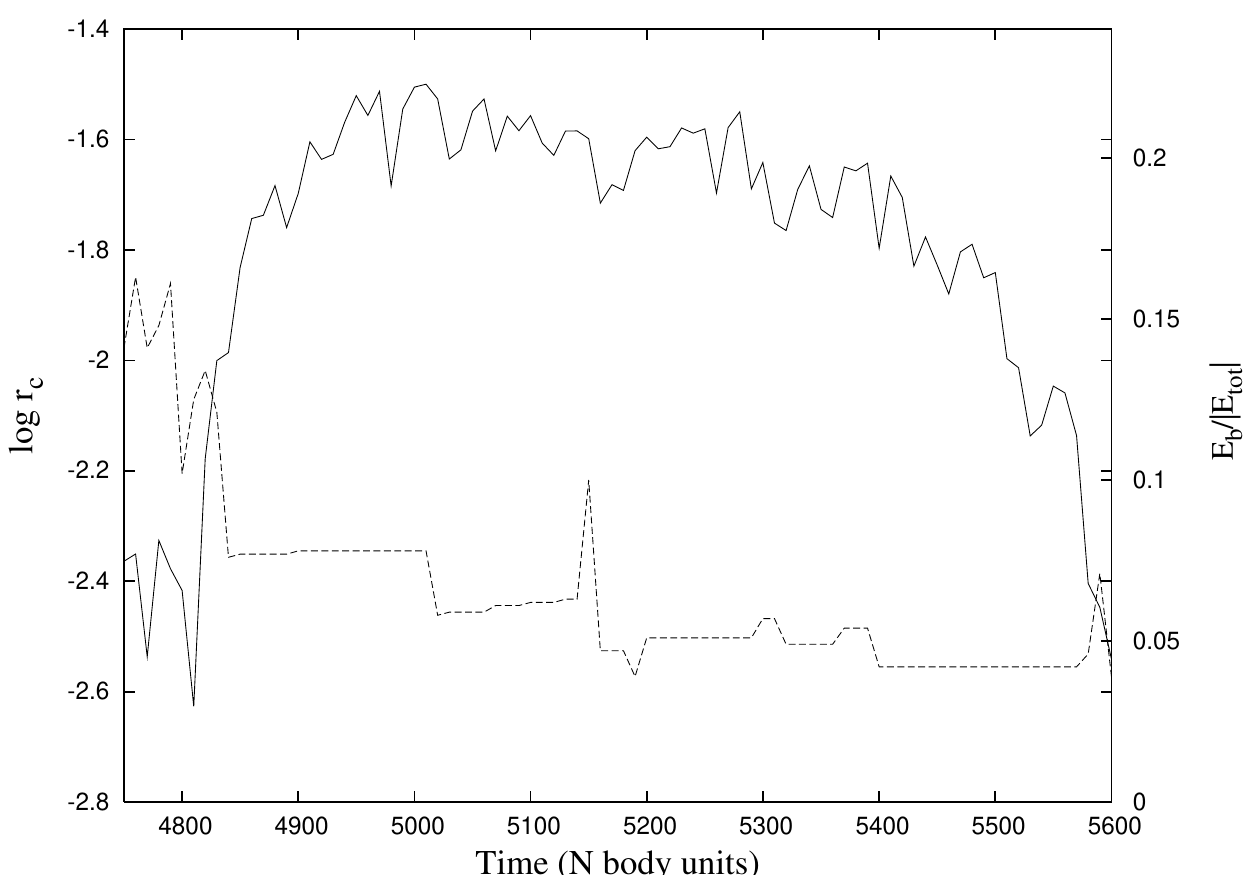}}}\quad
\subfigure{\scalebox{0.65}{\includegraphics{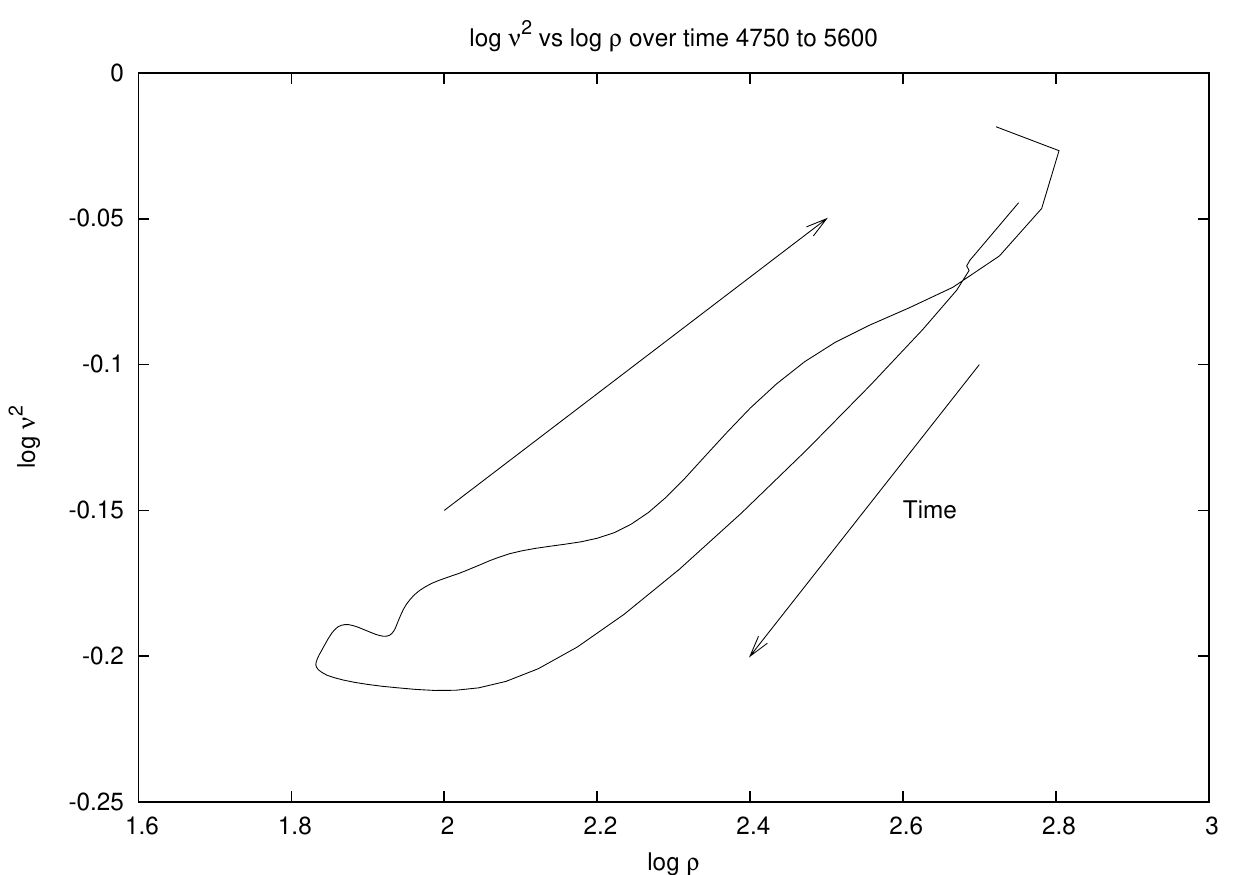}}}
\caption{ As Fig. \ref{fig:32ka0m1}, with $(m_{max},m_{min})=(1.0,0.1)$, but $N=64k$  and $\alpha = -1.3$. The velocity dispersion has been smoothed to make the cycle clearer.}
\label{fig:64ka135m1}
\end{figure}

For the case $\alpha = -1.3$ and stellar mass range $(1.0,0.1)$ the value of $N_{crit}$ is $3.5 \times 10^4$ (see Table \ref{table:tab1}). No gravothermal behaviour was seen in the $N$-body runs with $N =$ $8k$, $12k$ and $32k$ for this set of conditions. The first signs of gravothermal behaviour occurred in the $64k$ run as would be expected from the above value of $N_{crit}$ obtained from SPEDI. The same three graphs shown for both the $\alpha = 0.0$ cases (see previous paragraph) are plotted for the $64k$ run in Fig \ref{fig:64ka135m1}.

In the graphs of $r_c$ for all cases (see Fig. \ref{fig:32ka0m1} top, Fig. \ref{fig:32ka0m2} top and Fig. \ref{fig:64ka135m1} top), behaviour can be seen which is qualitatively similar to gravothermal oscillations (see \cite{Makino1996}, \cite{TakahashiSTOC},  \cite{HeggieGiersz} and \cite{breenheggie1}). For the case of $\alpha=0.0$ with the mass range $(1.0,0.1)$ (Fig. \ref{fig:32ka0m1} top) one oscillation can be seen between $3830$ and $4570$, and another between $5970$ and $6560$. During each of these oscillations $r_c$ changes by more than a factor of $10$. Similarly for the case of $\alpha=0.0$ with the mass range $(2.0,0.1)$ (Fig. \ref{fig:32ka0m2} top) an oscillation in $r_c$ can be seen between $6800$ and $7950$, and part of an oscillation can also be observed after $8610$. In this model the change in $r_c$ is about a factor of $10$. Finally for the case of $\alpha=-1.3$ with the mass range $(1.0,0.1)$ (Fig. \ref{fig:64ka135m1} top) an oscillation in $r_c$ can be seen between $4800$ and $5600$, and part of an oscillation can also be observed after $7400$. The change in $r_c$ is about a factor of $10$, which is similar to the change in $r_c$ for the $\alpha=-1.3$ runs.

Now we consider the physical nature of these oscillations, which could in principle be driven by sustained binary activity or by gravothermal behaviour. A sign of gravothermal behaviour is that the binding energy of the binaries remains roughly constant during times of expansion \citep{McMillan}. This is because the expansion phase of a gravothermal oscillation should be driven by the core absorbing heat from the rest of the cluster rather than by energy generation. At core bounce, where $\rho_c$ reaches a local maximum, there is an increase in binary activity, and enough energy is produced to halt and reverse the collapse. This behaviour is particularly clear in Fig. \ref{fig:64ka135m1} (middle) where there is an initial increase in relative binding energy of the binaries coinciding with core bounce and the initial expansion. A binary escapes, and then there is a period of expansion during which the relative binding energy of the binaries remains nearly constant (from $4850$ to $5000$). Towards the end of the oscillation there is renewed binary activity corresponding to the next core bounce. There is also binary activity at other times during the oscillation, but it has no discernible effect on the evolution of $r_c$. Fig. \ref{fig:32ka0m1} (middle) is also a good example of gravothermal behaviour. Mild binary activity continues after core bounce (from $t = 3860$ to $t = 3930$) but expansion continues thereafter for a period. However in Fig. \ref{fig:32ka0m2} evidence of gravothermal behaviour is more ambiguous. We will discuss the case of Fig. \ref{fig:32ka0m2} in detail in the last paragraph of this section.

The cycles of $\rho_c$ vs the core velocity dispersion $v^2_c$, as seen in Fig. \ref{fig:32ka0m1} (bottom), Fig. \ref{fig:32ka0m2} (bottom) and Fig. \ref{fig:64ka135m1} (bottom), are believed to be a sign of gravothermal behaviour \citep{Makino1996}. During these cycles, the temperature is lower during the expansion where heat is absorbed and higher during the collapse where heat is released. The velocity dispersion in Fig \ref{fig:32ka0m2} (bottom) and Fig. \ref{fig:64ka135m1} (bottom) has been smoothed to make the cycle clearer. These cycles are similar to the cycles found by \cite{Makino1996} for one-component models and by \cite{breenheggie1} for two-component models.

The gravothermal nature of the behaviour is clearer in Fig. \ref{fig:32ka0m1}  (where $(m_{max},m_{min})=(1.0,0.1)$) than in Fig. \ref{fig:32ka0m2}  (where $ (m_{max},m_{min})=(2.0,0.1)$). This is perhaps surprising as both cases have the same value of $N_{crit}$ as found with SPEDI. We will now discuss a number of possible reasons for this apparent difference in behaviour. Firstly, as the values of $N_{crit}$ are only correct to $10\%$, it is possible that in reality the values could differ by up to $4 \times 10^{3} $.  Secondly, another issue is that in the gas model the mass function is discretised, resulting in a difference between the mass of the heavy component ($m_{10}$) and $m_{max}$. $m_{10}$ is about $14\%$ percent less than $m_{max}$ for the stellar mass range $(2.0,0.1)$ and $11\%$ for $(1.0,0.1)$. It is argued in Appendix \ref{A} that the stability of a system will increase if the average stellar mass inside the core is increased (while keeping the stellar masses outside the core approximately the same). This would imply that in both cases the values of $N_{crit}$ found with the 10-component models are underestimates for the onset of instability, and that the true value of $N_{crit}$ for $(2.0,0.1)$ is slightly higher than for $(1.0,0.1)$. Finally, the fact that the gravothermal nature of the behaviour is clearer in one run might simply be a stochastic effect.

For the purposes of this paper, we have not considered the evolution of multi-component systems in the regime $N<N_{crit}$. In Spitzer-unstable cases, there is no reason to doubt that this is characterised by mass-segregation, followed by post-collapse expansion powered by binary evolution as in the much smaller N-body models considered long ago by \cite{albada1967}, \cite{aarseth1968} and many more since.

\section{Summary and Discussion}\label{sec:conanddis}

The focus of this paper has been on the conditions for the onset of gravothermal oscillations in multi-component systems. We have investigated power law IMFs with different exponents and three different stellar mass ranges $(3.0,0.1)$, $(2.0,0.1)$ and $(1.0,0.1)$. A multi-component gas code has been used to obtain the values of $N_{crit}$. In order to verify the validity of the results direct $N$-body runs were carried out on appropriately chosen cases. The values of $N_{crit}$ found ranged from $2\times 10^4 $ to $10^5$, and varied with $\alpha$ and the stellar mass range. 

Motivated by \cite{Murphyetal1990}, who found that the total mass of the systems they studied could be used as an approximate stability condition, the value of $M_{crit}$ (the total mass of the system at $N_{crit}$) for each system was calculated (see Table \ref{table:mcrit}). While for a fixed mass range $M_{crit}$ does provide an approximate stability condition, the value of $M_{crit}$ varied by roughly a factor $m_{max}$. $M_{crit}$  can also be used as an approximate stability condition for the two-component models of \cite{breenheggie1} so long as the stellar mass ratio is fixed (see Appendix \ref{A}).

In order to find a more general stability condition we applied an extension of an idea first employed in \cite{breenheggie1}. They used a quantity called the effective particle number ($N_{ef}$). The value was useful because the two-component system that was being considered was expected to behave in roughly the same manner as a one-component system with $N_{ef}$ stars. In the present paper this idea has been extended to multi-component systems. The values of $N_{ef}$ for the multi-component models in this paper are given in Table \ref{table:mcritNef}. The variation in  Table \ref{table:mcritNef} is significantly less then that in either Table \ref{table:tab1} or Table \ref{table:mcrit}. A stability condition of $N_{ef} \sim 10^4$ covers most of the values of Table \ref{table:mcritNef} and indeed the two-component models of \cite{breenheggie1} (see Table \ref{table:aptab3}).

The Goodman Stability Parameter was also tested for the multi-component case (see Table \ref{table:goodmanepsilon1}). The critical values in Table \ref{table:goodmanepsilon1} were found to be lower than the value for a one-component model ($\log_{10} \epsilon = -2$) and also varied with $\alpha$ and, to a much lesser extent, with $m_{max}$. By modifying the Goodman Stability Parameter using a slightly different definition for the half-mass relaxation time (based on the effective particle number) a critical value was found which was more consistent with the critical value for a one-component model (see Table \ref{table:goodmanepsilon2}).

\cite{Goodman1987} used a gas model to find the value of $N_{crit}$ ($=7000$) for a single component system. Technically what he showed was that steady post-collapse expansion was possible in a gas model for all $N$, but that it was unstable for $N>7000$. While the gas model used by \cite{Goodman1987} is similar in form to the model used here and in \cite{breenheggie1}, there are two notable differences. Firstly \cite{Goodman1987} used a larger energy generation rate than the one used here. Secondly, the parameter of the coulomb logarithm that was used was $\lambda=0.4$. A value of $\lambda=0.4$ \citep{Spitzer} was a reasonable choice at the time, but it has since been shown that $\lambda=0.11$ is a better choice for a single-component model \citep{GierszHeggie1994}. (For multi-component models the value of $\lambda=0.02$ was found to provide a good fit  \citep{GierszHeggie2}). These two differences affect the stability in opposite ways$:$ by arguments similar to those given in the last two paragraphs in Appendix \ref{A}, a larger energy generation rate will increase stability, whereas a larger value of  $\lambda$ tends to reduce stability. For example for $N=7000$ with $\lambda=0.4$ the increase in the relaxation rate is $20\%$ compared with $\lambda=0.11$. 

In the present paper, we have made the assumption that multi-component systems will be depleted in stars with stellar mass greater than $3 M_{\sun}$. This neglects the possibility of systems containing a population of stellar mass Black Holes, which would require a value of $m_{max}$ about an order of magnitude greater than what is considered here. These systems are outside the parameter space studied by \cite{breenheggie1} and \cite{KimLeeGood1998}, as the total mass ratio is lower than the range considered by \cite{breenheggie1} and the stellar mass ratio is higher than the values considered by \cite{KimLeeGood1998}. The onset of gravothermal oscillations and the more general evolution of systems containing a population of stellar mass black holes are the topics of the next paper in this series.

To conclude, a stability condition of $N_{ef} \sim 10^4$ does apply to the multi-component systems in this paper and the two-component systems of \cite{breenheggie1}. This condition is expected to apply to any multi-component system provided that there is a sufficient number of stars with stellar mass $\sim m_{max}$.

\section*{Acknowledgments}
We are indebted to S. Aarseth and K. Nitadori for making publicly available their version of NBODY6 adapted for use with a GPU. We would like to acknowledge R. Spurzem for the use of SPEDI and P. Amaro-Seoane for providing us with his version of the code. Our hardware was purchased using a Small Project Grant awarded to DCH and Dr M. Ruffert (School of Mathematics) by the University of Edinburgh Development Trust, and we are most grateful for it. PGB is funded by the Science and Technology Facilities Council (STFC).

\appendix
\section{The two-component case revisited}\label{A}

The purpose of this appendix is to reconsider the results of \cite{breenheggie1} in terms of the effective particle number $N_{ef}$ defined in equation \ref{eq:Nef}. \cite{breenheggie1} investigated gravothermal oscillation in a range of two-component models, specified by the stellar mass ratio $\frac{m_2}{m_1}$ and total mass ratio $\frac{M_2}{M_1}$, where $m_2$ ($m_1$) is the stellar mass of the heavy (light) component and $M_2$ ($M_1$) is the total mass of the heavy (light) component. For reference the values of $N_{crit}$ for these models are given in Table \ref{table:aptab1}, which is similar to Table 2 in \cite{breenheggie1}. The difference is that the data have been rearranged to compare more closely to the arrangement in the present paper. Thus the columns in Table \ref{table:aptab1} are arranged in order of decreasing $\frac{M_2}{M_1}$ as this is the analog for two components of decreasing $\alpha$. 

\begin{table}
\begin{center}
\caption{Critical value of $N$ ($N_{crit}$) in units of $10^4$ }
\begin{tabular}{ c  |c c c c c c}
\bf{$\frac{m_{2}}{m_1}\backslash\frac{M_2}{M_1}$ }                &  \bf{1.0}   & \bf{0.5}   & \bf{0.4}   & \bf{0.3}   & \bf{0.2}   & \bf{0.1}  \\ \hline
\bf{50}         & 18   & 30  & 33  & 42 & 55 & 100  \\ 
 \bf{20}         & 8.5  & 13  & 15  & 18 & 22 & 36   \\
 \bf{10}         & 5.0  & 7.2 & 8.2 & 10 & 12 & 22 \\ 
 \bf{5}       & 2.8  & 4.0 & 4.6 &5.4 & 7.0 & 10 \\ 
 \bf{4}       & 2.4  & 3.5 & 3.8 & 4.6 & 5.5 & 8.5\\ 
 \bf{3}       & 2.0  & 2.8 & 3.2 & 3.6 & 4.4 & 6.0 \\
 \bf{2}       & 1.7  & 2.2 & 2.3 & 2.6 & 3.0 & 3.8 
\label{table:aptab1}
\end{tabular}
\end{center}
\end{table}

Following the approach in the main part of this paper, we will firstly consider the values of $M_{crit}$. In order to consider $M_{crit}$ we need to specify the mass unit, and to make the results comparable with the multi-component models in the main part of the present paper $m_1$ has been fixed at $0.1M_{\sun}$. The values of $M_{crit}$ for the two-component models in Table \ref{table:aptab1} are given in Table \ref{table:aptab2}. For comparison the value of $M_{crit}$ for a one-component model would be $0.07\times 10^4$ (using $m=0.1M{\sun}$). This is significantly lower than any of the values in Table \ref{table:aptab2}\footnote{In Table \ref{table:aptab2} for fixed $\frac{m_2}{m_1}$, $M_{crit}$ increases with decreasing $\frac{M_2}{M_1}$. Given that $M_{crit}$ for $\frac{M_2}{M_1}=0$ is significantly lower than any of the values in Table \ref{table:aptab1}, one may wonder if that increasing trend observed in Table \ref{table:aptab1} continues below $\frac{M_2}{M_1}=0.1$. This is a topic considered in detail in the next paper of this series.}. For fixed values of  $\frac{m_2}{m_1}$ the value of $M_{crit}$ varies by factors of up to $\approx 3$ between $\frac{M_2}{M_1}=1$ and $0.1$. Therefore for fixed $\frac{m_2}{m_1}$, $M_{crit}$ does provide a rough stability condition. However for fixed $\frac{M_2}{M_1}$, the variation in $M_{crit}$ is a factor of $\approx 15-30$. The variation of $M_{crit}$ with varying $m_2$ resembles the variation of $M_{crit}$ with varying $m_{max}$ in Table \ref{table:mcrit}.

\begin{table}
\begin{center}
\caption[Caption for LOF]%
{Critical value of $M_{crit}$ in units of $10^4M_{\sun}$. The value of $m_1$ is fixed at $0.1M{\sun}$. For reference the value of $M_{crit}$ is  $0.07\times 10^4M_{\sun}$ for$\frac{M_2}{M_1}=0$, which is obtained from the result of \cite{Goodman1987} for a one-component cluster with $m_1 = 0.1 M_{\sun}$.}
\begin{tabular}{ c  |  c c c c c c}
  \bf{$\frac{m_{2}}{m_1}\backslash\frac{M_2}{M_1}$ }                 &  \bf{1.0}   & \bf{0.5}   & \bf{0.4}   & \bf{0.3}   & \bf{0.2}   & \bf{0.1}  \\ \hline
 \bf{50}     &   3.529 & 4.455 & 4.583 & 5.427  & 6.574  & 10.978   \\ 
 \bf{20}     &   1.619 & 1.902 & 2.059 & 2.305  & 2.614  & 3.940   \\ 
 \bf{10}     &   0.909 & 1.029 & 1.104 & 1.262  & 1.412  & 2.396   \\
 \bf{5}   &   0.467 & 0.545 & 0.596 & 0.662  & 0.808  & 1.078   \\ 
 \bf{4}   &   0.384 & 0.467 & 0.484 & 0.556  & 0.629  & 0.912   \\ 
 \bf{3}   &   0.300 & 0.360 & 0.395 & 0.425  & 0.495  & 0.639   \\ 
 \bf{2}   &   0.227 & 0.264 & 0.268 & 0.294  & 0.327  & 0.398    
\label{table:aptab2}
\end{tabular}
\end{center}
\end{table}

We will now consider the values of $N_{ef}$ for the two-component systems; these are given in Table \ref{table:aptab3}. For comparison the critical value of $N_{ef}$ for a one-component model is the same as its value of $N_{crit}$, which is $0.7\times10^4$. The values in Table \ref{table:aptab3} vary much less then those of $M_{crit}$ in Table \ref{table:aptab2}, although, as pointed out in Section \ref{sec:Ncrit}, $N_{ef}$ can be interpreted as a measure of the total mass of the system in units of $m_2$. A stability condition of $N_{ef} \sim 10^4$ or slightly more covers, within a factor $2$ at most, the values of Table \ref{table:aptab3} and indeed Table \ref{table:mcritNef}. 



\begin{table}
\begin{center}
\caption{Critical value of $N_{ef}$ in units of $10^4$ }
\begin{tabular}{ c  |c c c c c c}
\bf{$\frac{m_{2}}{m_1}\backslash\frac{M_2}{M_1}$ }                         &  \bf{1.0}   & \bf{0.5}   & \bf{0.4}   & \bf{0.3}   & \bf{0.2}   & \bf{0.1}  \\ \hline
 \bf{50}         & 0.71  & 0.89 & 0.91  & 1.09  & 1.31  & 2.20   \\ 
 \bf{20}         & 0.81  & 0.95 & 1.02  & 1.15  & 1.31  & 1.97   \\ 
 \bf{10}         & 0.91  & 1.03 & 1.10  & 1.26  & 1.42  & 2.40   \\ 
 \bf{5}       & 0.93  & 1.09 & 1.19  & 1.32  & 1.62  & 2.16   \\ 
 \bf{4}       & 0.96  & 1.17 & 1.21  & 1.39  & 1.57  & 2.28   \\ 
 \bf{3}       & 1.00  & 1.20 & 1.32  & 1.42  & 1.65  & 2.13   \\ 
 \bf{2}       & 1.13  & 1.32 & 1.34  & 1.47  & 1.64  & 1.99    
\label{table:aptab3}
\end{tabular}
\end{center}
\end{table}

All of the trends in Table \ref{table:aptab3} may be understood if we consider the reasoning behind the use of $N_{ef}$ as an approximate stability condition. The basic idea is that the multi-component system in question evolves in a similar way to a one-component system of $N_{ef}$ stars with stellar mass $m_2$. This requires that the half mass relaxation timescale of the multi-component system is similar to that of the one-component system with which we are comparing it. We assume this to be true if the heavy component amounts to a significant faction of the total mass within the half-mass radius $r_{h}$, which is certainly not the case as $\frac{M_2}{M_1}$ tends to $0$, i.e. on the extreme right of Table \ref{table:aptab3}.

We now consider with more care how the two-component system actually differs from the corresponding one-component system as the parameters $\frac{M_2}{M_1}$ and $\frac{m_2}{m_1}$ are varied. For fixed $\frac{m_2}{m_1}$, as $\frac{M_2}{M_1}$ decreases the relaxation process is increasingly dominated by the light stars. This leads to the system behaving more like a one-component system of $\frac{M_{tot}}{m_1}$ stars as opposed to a one-component system of $N_{ef}$ stars, and this increases the half-mass relaxation time. As the rate of two body relaxation becomes slower the core becomes larger (relative to $r_h;$ see the discussion of  H\'{e}non's Principle in Section \ref{sec:lsm}) as it can produce the required energy at a lower mass density (as the average stellar mass in the core remains the same, roughly $m_2$). Because gravothermal instability depends on a high density contrast within the system, it would be expected that stability would increase as $\frac{M_2}{M_1}$ decreases, as can be seen in Table \ref{table:aptab3}. 

Now let us consider the case of fixed $\frac{M_2}{M_1}$. If we consider the post collapse evolution of series of systems with fixed $M_{2}$ and $m_2 $, as $\frac{m_2}{m_1}$ decreases the tendency towards mass segregation becomes weaker. Therefore the half mass radius of the heavy component ($r_{h,2}$) is smaller compared to $r_{h}$ for larger $\frac{m_2}{m_1}$ than for smaller $\frac{m_2}{m_1}$. It follows that the mass density of the heavy component (within $r_{h,2}$) is smaller for smaller $\frac{m_2}{m_1}$ than for larger $\frac{m_2}{m_1}$. The relaxation time of the heavy component within its half-mass radius $r_{h,2}$ ($t_{rh,2}$) decreases with increasing mass density. This leads to the conclusion that the relaxation time within the heavy component increases with decreasing $\frac{m_2}{m_1}$. The energy flux in the heavy component, which we are assuming regulates the rate of energy generation, is of order $\frac{|E_2|}{t_{rh,2}}$ (where $E_2$ is the energy of the heavy system). Therefore, as $\frac{m_2}{m_1}$ decreases so does the energy flux, which results in a lower rate of energy generation. The lower rate of energy generation leads to a larger core (relative to $r_h$) as the core can produce the required energy at a lower mass density. Thus it would be expected that stability (as measured by $N_{ef}$) would increase as $\frac{m_2}{m_1}$ decreases, and this is what is observed in Table \ref{table:aptab3} for most values of $\frac{M_2}{M_1}$. However, the trend of increasing stability with decreasing $\frac{m_2}{m_1}$ seems to disappear for small $\frac{M_2}{M_1}$. Reasons for this will be discussed in the next paper of this series.

\bsp

\label{lastpage}

\end{document}